\documentclass[prb,aps,amssymb,showpacs,twocolumn,amsmath,floatfix]{revtex4}
\usepackage{tabularx}
\usepackage{bm}
\usepackage{euscript}
\usepackage{epsfig,psfrag,subfigure}
\usepackage{graphicx}
\usepackage{color}
\usepackage{amsfonts}
\usepackage{exscale}
\usepackage{amsbsy}
\usepackage{hyperref}
\input{epsf}

\def\be{\begin{equation}}
\def\ee{\end{equation}}
\def\ba{\begin{eqnarray}}
\def\ea{\end{eqnarray}}

\newcommand{\beq}{\begin{equation}}
\newcommand{\eeq}{\end{equation}}
\newcommand{\bea}{\begin{eqnarray}}
\newcommand{\eea}{\end{eqnarray}}

\begin{document}

\title{Effects of longer-range interactions on unconventional superconductivity  }
\author{S. Raghu$^1$, E. Berg$^2$, A.V. Chubukov$^3$ and S. A.Kivelson$^1$}
\affiliation{$^1$Department of Physics, Stanford University, Stanford, CA 94305}
\affiliation{$^2$Department of Physics, Harvard University, Cambridge, MA 02138}
\affiliation{$^3$Department of Physics, University of Wisconsin, Madison, WI 53706}
\date{\today}

\begin{abstract}
We analyze the effect of the non-vanishing range of
electron-electron repulsion on the mechanism of unconventional
superconductivity.
 We present asymptotically exact
 weak-coupling results
 for dilute electrons in the continuum and for the 2D extended Hubbard model, as well as
 density-matrix renormalization group results for the two-leg extended Hubbard model at intermediate couplings, and approximate results for the case of realistically screened Coulomb interactions.
 We show that  $T_c$ is generally suppressed in some pairing channels as longer range interactions increase in strength,
but superconductivity is not
 destroyed.
Our results confirm that electron-electron interaction
 can lead to unconventional superconductivity under physically realistic circumstances.
  \end{abstract}


\maketitle

\section{Introduction}

That unconventional superconductivity can arise in models with
short-ranged repulsive interactions between fermions has been
known for some time, beginning with the pioneering work by Kohn
and Luttinger~\cite{Kohn-Luttinger}. This issue has been revisited
multiple times in the last few decades, since the discovery of
unconventional superconductivity in the cuprates.  For the Hubbard
model with local repulsion $U$ and fermionic bandwidth $W$, the
existence of  superconductivity in $p$, $d$, $f$ or $g-$wave, as
well as sign-changing $s-$wave
   channels has been established
   from asymptotic weak coupling analysis  in the limit $U/W \ll 1$ in two and three
   dimensions~\cite{Fay-Layzer,Chubukov-Kagan,Baranov-Kagan,Hlubina,Chubukov-Lu,Chubukov,Raghu2010}, from the observation of a positive pair-binding
   energy computed from exact diagonalization on various ``Hubbard molecules,''~\cite{hubbard_molecules}
   from dynamical cluster approximation (DCA)~\cite{Maier} and    cluster dynamical mean-field theory (DMFT) calculations~\cite{Haule},
   and from extensive density matrix renormalization group (DMRG) studies\cite{DMRG_ladders,checkerboard} of various ladder systems extrapolated to the thermodynamic
    limit.

While there is still some controversy over the strength of the pairing tendencies under particular circumstances, there is a growing consensus
that superconductivity in the repulsive Hubbard model is generic under a wide range of circumstances.
 Various physically motivated approximate calculations, including
numerically implemented  functional renormalization group (FRG)\cite{FRG},
 dynamical cluster approximation (DCA)~\cite{Maier} and    cluster dynamical mean-field theory (DMFT) calculations~\cite{Haule},
  fluctuation exchange approximation
 (FLEX)~\cite{Flex}, Eliashberg~\cite{acn}
 and self-consistent two-particle calculations~\cite{Kyung},
as well as strong coupling approaches based on variational wave-functions~\cite{variational} and slave-particle mean-field theories\cite{slaveboson,lee_review}, have also
 strongly indicated that such unconventional pairing is present, especially near half-filling, and that $T_c$ is maximized in the
  physically relevant range of intermediate coupling, $U \sim W$, and decreases at both larger and smaller $U$.  The decrease of  $T_c$ at smaller $U$ is due to the fact that the strength of any induced attractive interaction
 must vanish as $U\to 0$, while the decrease
 as $U\to\infty$ is due to Mott physics which tends to localize fermions near particular lattice
   sites~\cite{steve_review,lee_review}, potentially suppressing both the superfluid stiffness\cite{EK} and the pairing tendencies\cite{infiniteU} of the electrons.

It is  broadly (although not universally\cite{imada}) accepted that the basic features of superconductivity arising from short-range repulsion
between electrons are
moderately generic and
must be in play in a broad class of unconventional superconductors such as the cuprates, heavy fermion and organic
superconductors, Sr$_2$RuO$_4$,
 and the recently discovered Fe-pnictides, despite differences in band-structure and local quantum chemistry.~\cite{basov}
 In particular,
 estimates of the optimal $T_c$ for d-wave pairing in 2D Hubbard models,
 obtained using a variety of
approximate   computational approaches~\cite{Maier,Haule,DMRG,Flex,acn,Kyung,variational,checkerboard},
suggest that
 ${\rm Max}[T_c]
\sim 10^{-2} v_F/a_0$ where $v_F$ is the Fermi velocity averaged
over the Fermi surface (FS), and $a_0$ is the interatomic
distance. Using $v_F/a \sim 1eV$,~\cite{vF_cuprates} one obtains
an estimate of the optimal $T_c$ of order 100K, comparable to the
d-wave transition temperatures found in optimally doped cuprates.

  However, in addition to the vexing problem of how to unambiguously establish or falsify the applicability of a particular electronic
  pairing mechanism to real materials,
  at least one key theoretical question remains to be addressed:
  It is well known that longer-ranged components of the electron-electron interaction, even simply a nearest-neighbor repulsion $V$ between electrons,
   suppress
   the pairing tendencies of the Hubbard model.  As we will discuss below, this can be seen clearly from the structure
    of the asymptotic weak-coupling approach and also  in exact diagonalization studies of Hubbard molecules.
   This effect has also been investigated in DMRG studies of t-J ladders\cite{Gazza1999} (which we extend to Hubbard ladders in the present paper).
   Thus, the issue to be addressed
    is whether the deleterious effects of longer-range components of the electron-electron repulsion make a purely electronic mechanism of superconducting pairing physically implausible.

The physics behind the suppression of $T_c$ is
transparent:  an effective attraction
which can give rise
 to unconventional superconductivity in $p-$wave, $d-$wave, extended $s-$wave, and other channels
   emerges in the theory
   from the renormalization (screening) of the local repulsive interaction $U$ by
  particle-hole fluctuations of the continuum of fermions.
  For small $U$, the induced interactions are necessarily of order $U^2/W$ or smaller, and more generally one expects the induced interactions to be overall weaker than the bare interactions that give rise to them.  However,
  if the bare interaction is short-ranged, then it only contributes in the trivial s-wave channel, while
  because the polarization  of the particle-hole continuum depends on the
  energy and
   momentum transfer between initial and scattered fermions, the
   induced interactions are non-local and may have attractive contributions in
   other pairing channels.

If, however, the bare interaction is extended, so it contributes a repulsive piece in
 the same
  extended $s-$wave,  $p-$wave, $d-$wave, etc, channels, the competition between the bare and induced interactions becomes more serious.
 This issue has been recently
  addressed
   in a particularly pointed fashion by Alexandrov and Kabanov (AK) (Ref.~\cite{alexandrov}).
   They have concluded that in models with a   screened Coulomb interaction, the bare repulsion    in non-s-wave channels is so significant that it effectively  overwhelms the  induced attractions, thus eliminating as physically realistic  the entire class of theories
 in which the
  pairing comes from electron-electron
  repulsion.

We will show that the conclusion of Ref.~\cite{alexandrov} is not warranted.
 From exact diagonalization and DMRG studies discussed below, it is clear that there is no anomalous sensitivity of the pairing strength to longer-range interactions;  in an extended Hubbard model, significant suppression of unconventional pairing occurs only when the farther neighbor interactions, $V$, are a noticeable fraction of $U$, and not when they exceed the (much smaller) gap scale associated with pairing.  From the various approximate extensions of the weak-coupling analysis to intermediate coupling, it is clear that a key physical ingredient missed in the discussion of AK is the near resonant enhancement of the induced attractions at special favored wave-vectors when the system is in (not too close) proximity of a density wave ordering transition;  under these circumstance, if the favored vectors appropriately nest the FS, the appropriate weighted average of the bare and induced interactions can favor unconventional superconductivity, despite the fact that the induced interactions are generically weaker than the bare interactions.~\cite{Dzyaloshinski,anderson-morel,d-wave}


 Even in the strictly weak coupling limit, where controlled calculations can be carried through (as shown below), the effect of longer range interactions turns out to be surprisingly muted.
 Longer-ranged interactions in lattice models
 (interactions between nearest neighbors, second-nearest neighbors, etc)
  generally have components only in some particular
   pairing channels and do not contribute to other channels. On the other hand, the
   particle-hole polarization bubbles  which enter the induced interactions
   have
   components in all channels, and
    quite often more than one component is attractive. In this situation, it is quite possible that
    longer-range interaction either does not affect
    the leading pairing component at all, or it suppresses the leading attractive component but
     leaves the subleading one unaffected.

In this context, we revisit in this paper the issue of unconventional superconductivity from electron-electron repulsion in
 an extended Hubbard model on a square lattice with an onsite repulsion of magnitude $U$ and nearest-neighbor and next-nearest-neighbor
   couplings of  magnitudes $V$ and $V'$, respectively. We study this problem using the same 
   asymptotic weak-coupling methods that were previously
applied to the pure Hubbard model, complemented by a new DMRG study of a two-leg ladder. We consider arbitrary fermionic density and analyze the pairing problem for various
   relations between $V, V'$, $U$, and the bandwidth $W$.   It is important to stress that the physics governing  $U$ and $V$ is generally
  quite distinct, and they should be considered as essentially independent parameters.
    For instance, in a transition metal oxide, $U$ reflects the atomic physics at short distances,
     while $V$ reflects the screening effects of all the degrees of freedom that are integrated out at high energies,
     including  the polarization of the surrounding medium in the solid state environment.
 We find that the interplay between the short and longer-range repulsive interactions results in a fascinating,
  complex phase diagram with a large number of distinct unconventional superconducting phases.
    The longer-range interactions do indeed tend to suppress
   unconventional superconductivity in some channels, but still we find that
     unconventional superconductivity emerges for all densities and for all relations between parameters of the model.

 This paper is organized as follows.  In the next section, we review the weak coupling approach to pairing in a system of fermions with 
generic
finite-range interactions.  In Sec. \ref{sec:isotropic} we discuss the case of
 short-range interactions and
 small electron density,
where the fermionic dispersion can be
approximated by a parabola. In \ref{sec:hubbard} we extend the approach to arbitrary electron densities and
 analyze the extended Hubbard model with nearest, and next-nearest neighbor interactions.
  In Sec. \ref{sec:coulomb} we consider the case of screened Coulomb interactions, which we can only treat approximately,
  even in the small $r_s$ limit.
  In section \ref{sec:ladders}, we present the results of the DMRG  calculations of the pairing amplitudes in Hubbard ladders.
  We discuss the results and present our conclusions in Sec.\ref{sec:discussion}.

\section{Asymptotically exact weak-coupling approach}
\label{sec:general}

The  basic idea of the Kohn-Luttinger (KL) approach to superconductivity is that the pairing interaction is given by the irreducible vertex in the
  particle-particle channel, which is generally different
 from the bare interaction between the two given fermions and includes contributions from the continuum of particle-hole excitations.
  These additional contributions to the pairing interaction can give rise to an attraction in a particular pairing channel even if the original
   interaction between two fermions is
   uniformly repulsive.

 The KL approach
 can be formally justified if
 there is a separation of scales:  superconductivity is assumed to come from fermionic
 states very near the FS, while the renormalization of the irreducible pairing interaction comes from fermions with energies
  comparable to the Fermi energy.  In this situation the fully renormalized pairing interaction can be approximated by
   its value when both incoming and outgoing fermions are on the FS and the two incoming fermions have opposite momenta and zero energy
    (where the energy is measured as deviation from the chemical potential).
    Such a separation of scales
    clearly occurs when the interaction $U$ is much smaller than the fermionic bandwidth $W$;
    under appropriate circumstances, it may hold approximately (in a physical sense) even when
   $U$ is comparable to the bandwidth~\cite{acn}.

 The KL approach
 can be implemented in terms of
a two-step renormalization group (RG)
 procedure~\cite{Raghu2010}:
In the first step, we integrate out all modes outside a narrow
range of energies $\Omega_0$ about the Fermi energy. $\Omega_0$ is
not a physical energy in the problem, but rather a calculational
device. It is chosen large enough so that the
 interactions can be treated perturbatively (e.g., the Cooper logarithms are still small),
  but small enough that it can be set to zero in all renormalizations in non-singular channels
  without causing significant error, {\it i.e.} it is chosen to satisfy the inequalities
 \be
 N_F U^2 \gg \Omega_0 \gg E_F\exp\{-[1/(N_F U)] \},
 \label{Lambda}
 \ee
where $N_F$ is the density of states (DOS) at the Fermi energy (for both spin orientations) and $E_F$ is the Fermi energy.
 The effective interactions generated in the process then serve as input interactions in a second step,
 in which the remaining problem is solved using the logarithmical RG technique~\cite{Shankar1994,Polchinski1992} which is equivalent to BCS when only
 the pairing channels
 are singular. (It  involves a more complex parquet RG if some  other channel,
 e.g., SDW or CDW,
  is also singular and competes with
 superconductivity,~\cite{Polchinski1992,Dzyaloshinski}
 a situation we will not consider explicitly for the present.)
  $T_c$ is, up to a  multiplicative constant, given by the energy scale,
  $T^*$, at which the pairing interaction grows to be of order
   one.
     It was shown by explicit perturbative calculation of $T_c$ 
   up to 4th order in $U$ that the resulting expression for $T^*$ is independent of $\Omega_0$ (Ref. \cite{Raghu2010,maxim_fullTc}).
  \begin{figure}
\includegraphics[width=3.0 in]{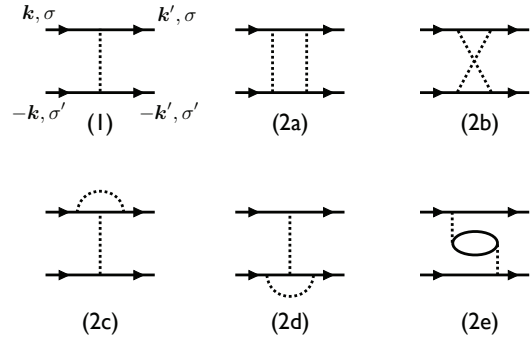}
\caption{First and second order diagrams which contribute to the effective interaction $\Gamma^{a}({\bf k}, {\bf p})$ in the Cooper channel.
  The solid line corresponds to an electron propagator whereas the dashed line represents an interaction vertex, $\sigma$ and $\sigma'$ are spin
  indices. The singlet and triplet components $\Gamma^{(s)}$ and $\Gamma^{(t)}$  are obtained by antisymmetrizing the interaction and taking
  symmetric and antisymmetric combinations, $(\Gamma ({\bf k}, {\bf p})+ \Gamma^{a}({\bf k}, -{\bf p}))/2$ and
  $(\Gamma ({\bf k}, {\bf p}) - \Gamma^{a}({\bf k}, -{\bf p}))/2$, respectively.}
 \label{diag}
\end{figure}
The generic  prescription for computing the leading order asymptotic
 behavior of $T_c$ for weak interactions is the following~\cite{Raghu2010}:
   first, compute the effective interaction in the Cooper channel at energy scale $\Omega_0$
   perturbatively, to second order in the bare interaction. The corresponding diagrams are  shown in Fig. \ref{diag}.
   We denote this effective interaction as $\Gamma^{(a)}({\bf k}, -{\bf k}; {\bf p}, -{\bf p}) = \Gamma^{(a)}({\bf k}, {\bf p})$, where
    ${\bf k}$ and ${\bf p}$ denote points on the FS and
    the superscript $a$ differentiates between
     whether the electron pair forms a spin singlet
    ($\Gamma^{(s)}$) or a spin triplet ($\Gamma^{(t)}$).  We then construct the related dimensionless matrix
\be
\gamma^{(a)} ({\bf k}, {\bf p}) \equiv {\bar N}_F \sqrt{{\bar v}_F/v_F(\bf k)}\Gamma^{(a)}({\bf k},{\bf p}) \sqrt{{\bar v}_F/v_F(\bf p)},
\ee
where $v_F({\bf k})$ is the magnitude of the Fermi velocity at a given point ${\bf k}$ on the FS, and
${\bar v}_F$ and ${\bar N}_F$ are the average Fermi velocity and average DOS at the FS. For a circular FS,
$v_F({\bf k}) = {\bar v}_F = v_F$ and ${\bar N}_F =N_F$, but the approach also holds for lattice models of fermionic dispersion.

Since $\gamma^{(a)} ({\bf k}, {\bf p})$ is a real, symmetric matrix,  it has a complete set of eigenstates $\phi_{{\bf k}}^{(a,l)}$
 and eigenvalues $\lambda^{(a,l)}$
\be
\sum_{\bf p}\gamma^{(a)} ({\bf k}, {\bf p}) \phi_{{\bf p}}^{(a,l)} =\lambda^{(a,l)}\phi_{{\bf k}}^{(a,l)}.
\label{su_2}
\ee
 Since all the systems we will consider are inversion symmetric,  the spin symmetry is implicitly determined by whether $l$ transforms under an irreducible representation that is even or odd under inversion, so we will henceforth leave it implicit, $(a,l) \to (l)$.
 Among all the possible solutions, we identify the most negative eigenvalue,
\be
\lambda \equiv {\rm Min}\left[  \lambda^
{(l)}\right], ~~\lambda < 0
\ee
Then,
\be
T_c \sim E_F \exp[-1/\vert \lambda \vert ].
\ee
Since the effective interaction $\Gamma$ is computed perturbatively assuming weak interactions, $\lambda$ can be expressed 
as a power series in
 the bare interaction. Assuming
   that the interaction
 is a local Hubbard $U$, we have
 \beq
 \lambda = - \vert u \vert \left[ 1  + a_1 \vert u \vert + a_2 u^2 + \cdots \right]
 \eeq
 for negative $U$  ($u = N_F U$), and
 \begin{equation}
 \lambda = -u^2 \left[ b_0 + b_1 u + b_2 u^2 + \cdots \right],
 \end{equation}
 for positive $U$.  $a_i$ and $b_i$ are non-universal constants which depend on dimensionality, crystalline point group, 
  electron concentration, and details of the banstructure.  The absence of a linear in $u$ term in $\lambda$ for $U>0$ is 
  a consequence of the fact that in a repulsive Hubbard model,
   an attractive interaction in any pairing channel
   can only appear due to a renormalization of the original interaction
   by the particle-hole continuum. As
   a  consequence, $T_c$ is a highly non-analytic function of $u$ in the limit $u \to 0$, with different functional dependencies for positive and negative $u$.

In the next two sections we apply this general procedure to systems with isotropic dispersion, and to systems with
 full lattice dispersion.

 \section{Continuum electrons with weak short-ranged interactions}
\label{sec:isotropic}

Dilute electrons on a lattice treated in the effective mass approximation are equivalent to the continuum problem in which the FS is a sphere (a circle in 2D)
 and the electronic dispersion
 has  the parabolic form $\epsilon_k = k^2/2m - \mu$.
  As a first example, we thus consider the low density limit of electrons with an interaction, $U(r)$
of finite range, $r_0$, which is independent of the electron density.
 There are thus two independent small parameters:  $k_Fr_0\ll 1$ and $k_Fa \ll 1 $ where $a \equiv m\tilde U(0)/4\pi \sim m U(r=0) r^3_0$ is the s-wave scattering length in the Born approximation, and $\tilde U$ is the Fourier transform of $U$.

 To first order in $U$, the singlet and triplet components of the pairing interaction in the momentum space
 are
 \bea
 &&\Gamma^{(s)} ({\bf k}, {\bf p}) = 
 \left[\tilde U({\bf k}-{\bf p}) + \tilde U({\bf k}+{\bf p})\right]/2 \nonumber \\
 &&\Gamma^{(t)} ({\bf k}, {\bf p}) = 
 \left[\tilde U({\bf k}-{\bf p}) -\tilde U({\bf k}+{\bf p})\right]/2
 \label{su_4}
 \eea
  For ${\bf k}$ and ${\bf p}$ located at the FS, $({\bf k-p})^2 = 2k^2_F (1 - \cos \theta)$.
  Since $\tilde U({\bf k}
  )$  changes slowly
  over the range of
 momenta
 $0\leq |{\bf k}| \leq 2k_F$, it 
 can be expanded in powers of $k_F r_0 \ll 1$: 
\beq
\tilde U({\bf k}
) = \tilde U (0) \left(1 + \alpha_1 |{\bf k
}|^2 r^2_0 + \alpha_2 |{\bf k
}|^4 r^4_0 + ...\right)
\label{su_3}
\eeq
where $\alpha_n$ are pure numbers related to moments of $U(r)$.
The first term in (\ref{su_3}) then contributes to $s-$wave pairing, the second term contributes to $s-$wave and
$p-$wave pairing, the third term contributes to $s-$wave, $p-$wave, and $d-$wave pairing, and so on.
 Indeed, because there is full rotational symmetry, the pairing wave functions ({\it i.e.} the eigenstates in Eq. \ref{su_2}) are completely determined by symmetry ({\it i.e.} they are uniquely specified by
angular momentum sectors labeled by $l$).
 Correspondingly, the eigenvalues, $\lambda^{(l)}$,  can be computed directly as the appropriate symmetry determined average of $\gamma$ over the FS.
In 3D, $N_F = m k_F/(2\pi^2)$, and
 it is
important to note that
 $u = N_F {\tilde U}(0) =2 (ak_F)/\pi$ .

Then, following the renormalization group procedure outlined
above, explicit perturbative expressions can be obtained for
$\lambda^{(l)}$ which, to second order in $u$, are \be
\lambda^{(l)}=  (k_Fa) A^{(l)}+ (k_Fa)^2 B^{(l)}+ \ldots \ee where
$A^{(l)} $ and $ B^{(l)}$ are dimensionless functions of the
dimensionless variables, $k_F r_0$ and $N_F\Omega_0$ (where
$\Omega_0$ is the scale at which the couplings are defined, as
discussed above Eq. \ref{Lambda}).  However, we are interested
only in their leading behavior for small values of both these
parameters. Starting with the first order terms, $A^{(l)}$ is
manifestly independent of $N_F\Omega_0$ but varies as $k_Fr_0\to
0$ according to $A^{(l)} \sim (k_Fr_0)^{2l}$.  Specifically, for
the three lowest angular momentum channels, \bea &&A^{(s)}= \frac
{2}{\pi},~~A^{(p)}= -\frac {\alpha_1} {\pi} (r_0 k_F)^2,
\nonumber \\
&&A^{(d)}=\frac{ \alpha_2} {\pi} (r_0 k_F)^4\ .
\eea
The components with larger $l$
are progressively
smaller in powers of $r_0 k_F$.
 To this order, the sign of $\lambda^{(l)}$ depends on the sign of $a$ and on the sign of $\alpha_l$. For a repulsive  Yukawa potential ({\it i.e.} for $\tilde U(q) = \tilde U(0)/[1+(qr_0)^2]$),
 $a>0$, $\alpha_1 <0$ and $\alpha_2>0$, and this is rather generic for purely repulsive interactions.
Thus, there is no Cooper instability to first order in $(k_F a)$.

The situation is altered by corrections to $\Gamma^{(a)} ({\bf k}, {\bf p})$ to second order in $ak_F$, see Fig. \ref{diag}.
 The diagram 2a describes the renormalization in the particle-particle channel which makes a contribution to $B^{(s)}$ in the s-wave channel
 proportional to $\log[W/\Omega_0]$.  However, so long as the inequality in Eq. \ref{Lambda} is satisfied, this only produces a small correction to $\lambda^{(s)}$ relative to the first order repulsive term.
 The other four second-order diagrams which contribute to $B^{(l)}$ contain
    particle-hole bubbles and  account for the important renormalizations of the irreducible pairing interaction.

     A key result obtained already by KL~
     (Ref.\onlinecite{Kohn-Luttinger})
      is that,  as $k_F r_0\to 0$,
     $B^{(l)}\to \beta_l
     \neq 0$.
        KL
        further showed that for sufficiently large $l >0$, $B^{(l)}$ are negative, and that while $A^{(l)}$ falls exponentially with increasing $l$, $B^{(l)} \sim l^{-4}$.  From this, they concluded that superconductivity in some (possibly high $l$) channel was inevitable.

      We focus attention on the physically more important cases of relatively small $l$. It
      has been  shown ~\cite{Fay-Layzer,Chubukov-Kagan} that
      $\beta_l$ are
       negative for $l >0$.
         As a  result,
\bea
&&\pi \lambda^{
(p)} =(ak_F)\left[ |\alpha_1|  (r_0 k_F)^2 - |\beta_1| (ak_F)+ {\cal O}(k_Fa)^2\right] , \nonumber \\
&&\pi \lambda^{(
d)} = (ak_F)\left[|\alpha_2|  (r_0 k_F)^4 - |\beta_2| (ak_F)+  \ldots\right],
\label{ch1}
\eea
where all $\beta_l$ are 
of order 1. The issue then is which terms are larger, the negative second-order
or the positive first-order contributions.
Clearly for  $k_F^{-1} \gg r_0 \gg a$,
the repulsive part of the interaction
 is dominant.  However,
  there is a broad range of circumstances
    in which $k_F^{-1} \gg r_0 \sim a$.
      In this situation, the attractive interactions dominate, even for relatively small $l >0$.
      In 3D
  \beq
  \beta_1 =- \frac{8 (2 \log 2 -1)}{5\pi^2},
  ~~\beta_2 \sim 10^{-2} \beta_1
  \eeq
   (see Refs. \onlinecite{Fay-Layzer,Chubukov-Kagan}), and hence at small density a 3D Fermi system with short-range repulsive
   interactions undergoes a $p-$wave pairing transition with
   \beq
   T_c \propto \exp\left[-1/(|\beta_1| (ak_F)^2)\right].
   \eeq
    (For the full calculation of $T_c$ see Ref.\cite{maxim_fullTc}.) In 2D, the calculations are a bit more tricky because the static particle-hole bubble for free fermions is
   independent of momentum transfer
   for$|{\bf k} - {\bf p}| <2k_F$, so one has to go to the next, third order, to obtain the momentum dependence of the
    dressed interaction~
    (see Ref.\onlinecite{Chubukov}).
    The final result is, however, similar to the one in 3D:   a 2D Fermi system at a small density undergoes a $p-$wave superconductivity
     with a  low
   $T_c$.

   Note that a similar analysis was carried out in Ref~\onlinecite{kagan_V} for the extended Hubbard model at small electron densities, where instead of a Yukawa interaction, strong Hubbard interactions were considered with $U\gg V\gg $ the bandwidth with similar results to those discussed here.

 \section{Extended Hubbard model at weak coupling}
\label{sec:hubbard}

As a
second more directly experimentally relevant example,
 we apply the reasoning from Sec.\ref{sec:general} to a higher density of electrons in the extended Hubbard model on a two dimensional square  lattice with
 the
  Hamiltonian:
\begin{eqnarray}
H &=&H_0 + H_{int} \nonumber \\
H_0 & = & -t \sum_{\langle ij \rangle \sigma} c^{\dagger}_{i \sigma} c_{j \sigma} + H.c. \label{su_6} \\
H_{int} &=& U \sum_{i} n_{i \uparrow} n_{i \downarrow} + V \sum_{\langle i,j \rangle}  n_i n_j + V' \sum_{\langle \langle i,j \rangle \rangle}
 n_i n_j
 \nonumber \\
 \end{eqnarray}
where $c_{i \sigma} $ is the destruction operator of an electron on lattice site $i$, with spin $\sigma$, $\langle i,j \rangle$,  and $ \langle \langle i,j \rangle \rangle$ denotes  nearest-neighbor and second-neighbor pairs of sites respectively, 
 $n_i = \sum_{\sigma} c^{\dagger}_{i \sigma} c_{i \sigma}$ and the average density of electrons per site is $n\equiv <n_i>$.
 For simplicity, we consider band electrons with only nearest-neighbor hopping.  As is well-known, such a system possesses
  a non-generic particle-hole symmetry.  This feature, however, does not play a significant  role in the resulting phase diagram since we
  consider
  the model away from half-filling. (See, however, Ref. [\onlinecite{mraz_hlubina}]).

 To begin with, for $V' = 0$  we show that
  the phase diagram is drastically different depending
 on
  the ratio
  $V/U$, even as $u \sim U/t \to 0$.  Different asymptotic analysis is required for the case $V = \alpha U^2/W$
 ({\it i.e.} for $u \rightarrow 0$ with $VW/U^2 = \alpha > 0$ held constant), and $V = \alpha' U$ (i.e. for $u \rightarrow 0$ with
 $V/U = \alpha' > 0$ held fixed). We find that in both cases, the ground state is superconducting for all dopings, but the symmetry of the
 pairing state is generally different. We then  add $V' \sim V$ and show that it adds
  other
   pairing states to the phase diagram.

 \subsection{$V \sim U^2/W$, $V'=0$}

Since the effective interaction is computed to $\mathcal O(U^2)$, when $V = \alpha  U^2/W$, we only need
take
 $V$ into account to first order (Diagram 1 in Fig. \ref{diag}), i.e. at the bare level.  A non-zero V then produces a correction
to the effective $\Gamma
({\bf k}, {\bf p})$ in the form
\begin{eqnarray}
\delta \Gamma
({\bf k}, {\bf p}) &=& \alpha \frac{U^2}{W}  \left[ \cos{\left( k_x - p_x \right)} + \cos{ \left( k_y - p_y \right)} \right] \nonumber \\
&=& \alpha \frac{U^2}{W} \sum^\prime_{\left( \eta \right)}  \phi^{(\eta,1)}(\bf k)  \phi^{(\eta,1)}(\bf p)
\label{su_7}
\end{eqnarray}
where in the second line the sum is taken over appropriate basis functions defined on nearest-neighbor sites with   A$_{1g}$ or extended s-wave symmetry,
B$_{1g}$ or d$_{x^2-y^2}$-wave symmetry, and E$_u$ or p-wave symmetry:
\ba
&&A_{1g}: \ \ \ \ \ \ \ \phi^{(s,1)}({\bf k})=[\cos(k_x)+\cos(k_y)]/\sqrt{2},  \\
&&B_{2g}: \ \ \ \ \ \ \ \phi^{(x^2-y^2,1)}({\bf k})=[\cos(k_x)-\cos(k_y)]/\sqrt{2},\nonumber \\
&&E_u: \ \ \ \ \ \ \ \phi^{(x,1)}({\bf k})=\sin(k_x), \ \ \ \ \
\phi^{(y,1)}({\bf k})=\sin(k_y). \nonumber
\ea
 Thus, the nearest neighbor interaction acts as a separable repulsive interaction
 within these subspaces.
 Because all three components are repulsive, $V$ tends to suppress the pairing tendency in all three of these channels.  However, note that unlike the case in the continuum, there are multiple (infinite) orthogonal functions defined on the FS which transform according to each irreducible representation of the point group - for instance $\phi^{(s,2)}\sim [\cos(k_x)+\cos(k_y)]^2$ 
 also transforms according to A$_{1g}$  under operations of the point group.  Thus, even for large $\alpha$, the presence of a repulsive first order term in a given channel does not preclude the existence of a more ``extended'' form of pairing in the same channel.
  \begin{figure}
\includegraphics[width=3.5 in]{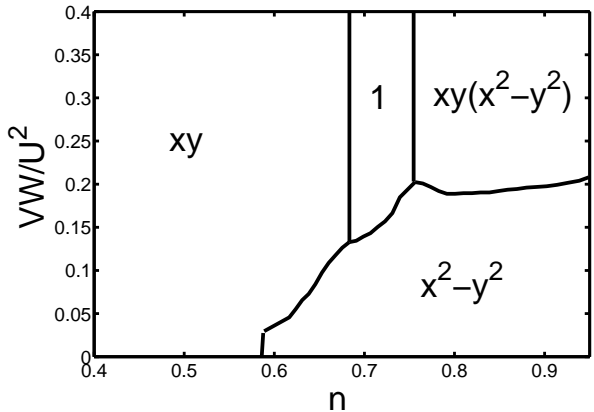}
\caption{Phase diagram as a function of electron concentration for the 2D extended Hubbard model in the regime where $V \sim U^2/W$ and $V'=0$.
 The region of extended $s-$wave superconductivity is labeled as $1$.
 For $V=0$, our phase diagram is similar but not identical to the one obtained by Hlubina~\cite{Hlubina}. In his calculation, there is a small region of $p-$wave superconductivity between $d_{xy}$ and $d_{x^2-y^2}$ phases. We found only d-wave states.}
\label{pd1}
\end{figure}
  \begin{figure}
\includegraphics[width=3.5 in]{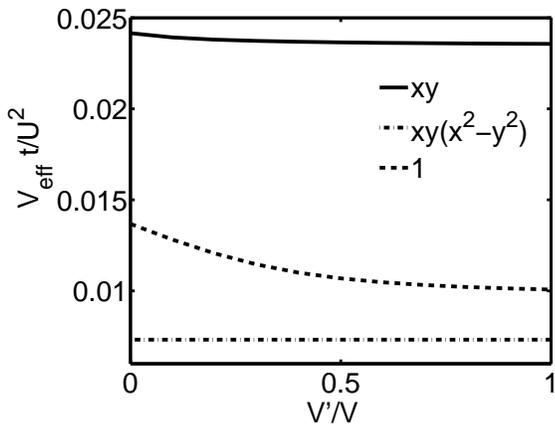}
\caption{The effective interaction, $V_{eff} = \lambda/N_F$ where
$N_F$ is the density of states at the Fermi level, as a function
of $V'/V = \alpha_2$.  $(n=0.6)$ with $V = \alpha U^2/W$ , $\alpha
= 0.16$ and non-zero $V'$.   } \label{pd2}
\end{figure}

It has been shown previously in various studies of the repulsive
$U$ Hubbard model  that near
half-filling,~~\cite{Hlubina,mraz_hlubina,Raghu2010,alexandrov}
 predominant pairing instability to order $U^2$ is in the
d$_{x^2-y^2}$
channel, and the subdominant pairing eigenvalue occurs in the A$_{2g}$ 
or g-wave channel, while somewhat further from half-filling,
 at $n < 0.62$,
  there is a range of electron concentrations for which the
 B$_{2g}$ or $d_{xy}$ pairing solution is dominant: Representative gap functions (of the shortest possible spatial range) with these symmetries are:
\ba
&&\phi^{(g,3)} \sim [\cos(k_x)-\cos(k_y)]\sin(k_x)\sin(k_y). \\
&&\phi^{(xy,2)} \sim \sin(k_x)\sin(k_y). \nonumber \ea

Figure \ref{pd1} shows the phase diagram to order $U^2$, which we obtained numerically  for $V = \alpha U^2/W$,
 as a function of $\alpha$ and $n$.
 Since the analysis considered here breaks down for concentrations
  sufficiently close to half-filling because of competition with antiferromagnetism,
   we have investigated solutions only for $0< n < 0.95$.  For $0.76 < n < 0.95$, a finite $\alpha \geq 0.2$
    destabilizes
   the $d_{x^2-y^2}$ solution in favor of the subdominant g-wave solution
   (labeled as $xy(x^2-y^2)$ on the phase diagram in Fig. \ref{pd1}).
   For $0.68 <  n < 0.76$ a finite $\alpha$ again destabilizes  the $d_{x^2-y^2}$ solution, but the state that emerges instead
   has
    a particular extended $s-$wave symmetry, that is dominantly of the form
     Re$(x+iy)^4$.  To shorten notations,
      have labeled it as ``1".
       This state has zero amplitude for on-site pairing in this limit but transforms nevertheless as a trivial irreducible representation of the tetragonal point group.  
       Note that since the $g-$wave state is Im$(x+iy)^4$, in the continuum limit, the ``1" and $g$ states are degenerate corresponding to angular momentum $\ell = 4$  pairing.  However, lattice effects lift this degeneracy.  Indeed, 
       there is a phase transition between $g-$wave and extended $s-$wave state at $n =0.76$,
   which reflects a level crossing of these two  eigenvalues, both of which are sub-leading at  $V=0$.
      Note that this phase boundary is vertical since the two solutions are
      unaffected by $V$ to first order.  For $0.58 < n < 0.68$, we have found that a finite $\alpha$ favors the d$_{xy}$ solution.
      The phase boundary between $d_{xy}$ and extended $s-$wave phases is also vertical.

 \subsection{$V \sim U^2/W$, $V' \sim V$}

Next, we consider the effect of a non-zero second neighbor interaction $V'$ on the phase diagram.
Specifically, we take $V' = \alpha_2 V, 0 < \alpha_2 < 1$.  In this case, $V' \sim U^2/W$, 
so like $V$, its effects can be computed via the first order
diagram 1 in Fig. \ref{diag}.  The expansion of the $V'$
interaction into angular harmonics is straightforward,
 and in addition to terms already present in Eq. (\ref{su_7}), one finds the following contribution to the effective Cooper channel interaction from $V'$:
 \begin{equation}
 \delta \Gamma_{V'}(\bm k, \bm p) = 2 \alpha \alpha_2 \frac{U^2}{W} \cos{\left( k_x - p_x \right)} \cos{\left( k_y - p_y \right)}
 \end{equation}
This term is a sum of separable repulsive interactions in the
extended-S (A$_{1g}$), d$_{xy}$ (B$_{2g}$), and p-wave (E$_u$)
subspaces.  It is important to stress that it does not affect the
d$_{x^2-y^2}$ (B$_{1g}$) or the g-wave (A$_{2g}$) subspaces. Thus,
relatively close to half-filling where the leading eigenvalues
belong to the d$_{x^2-y^2}$ and g-wave subspaces, the phase
boundaries are unaffected by $V'$.  We have studied the phase
diagram in the presence of non-zero $V'$ and have found that
while the phase boundaries are affected quantitatively by non-zero
$\alpha_2$, the topology of the phase diagram itself is unchanged.
We have observed that the pairing strengths are surprisingly
robust in this regime, as $\alpha_2$ is increased.
 Figure \ref{pd2} displays the effective interactions $V_{eff} = \vert \lambda \vert / N_F$, where $N_F$ is the density of states at the Fermi energy,  in the presence of non-zero $V'$.  For definiteness, we work in a fixed density of electrons $n=0.7$, and fixed $\alpha =0.16$ and show the effective interaction as a function of $\alpha_2$.  The d$_{x^2-y^2}$ pairing strength is the weakest one here and is not shown.  It is apparent from the figure that while the pairing strengths of the d$_{xy}$ and extended s-wave states are affected by non-zero $\alpha_2$, this repulsion is not strong enough
  to  suppress superconductivity altogether.

\subsection{$ V \sim U$}

Next, we consider the case when $V \sim U$, in which case the bare repulsive interactions are of the form
\begin{equation}
\Gamma_{I}(\bm k, \bm k') = U + Vg(\bm k - \bm k')
\end{equation}
where the subscript on $\Gamma$ above is to remind the reader that this comes from diagram I in Fig\ref{diag}.  The dimensionless function
\begin{equation}
g(\bm q )= \cos{q_x} + \cos{q_y} +2 \beta \cos{q_x} \cos{q_y}
\end{equation}
 specifies the momentum dependence of the interactions, where $\beta = V'/V$.
  We must now compute the perturbation expansion of the effective interaction in the Cooper channel to second order in all the bare interactions which has the form
  \begin{eqnarray}
 \Gamma^{(s)}(\bm k, \bm k') = &&\Gamma_{I}(\bm k, \bm k') + U^2 f_1^{(s)}(\bm k, \bm k') + V^2f_2^{(s)}(\bm k, \bm k') \nonumber \\
 && + UV f_3^{(s)}(\bm k, \bm k') + \cdots
 \end{eqnarray}
  where $s$ denotes the spin-singlet  channel.  It is straightforward to show that
  \begin{eqnarray}
  f_1^{(s)}(\bm k, \bm k') =
  N_F
    \ln{\left[ \frac{W}{\Omega_0} \right]}  + \chi(\bm k + \bm k') + \mathcal{O}(\Omega_0)
  \end{eqnarray}
  where $N_F$ is the density of states at the Fermi energy.
  The first term is obtained from diagram $2a$ and the second from $2b$ in Fig. \ref{diag}.  Similarly,
  \begin{eqnarray}
  && f_2^{(s)}(\bm k, \bm k')  =  N_F \left( g \star g\right) (\bm k - \bm k') \ln{\left[ \frac{W}{\Omega_0} \right]}  + \chi_{1}(\bm k, \bm k') + \nonumber \\
  &&   g(\bm k - \bm k') \chi_{2}(\bm k, \bm k') - 2\left[g(\bm k - \bm k') \right]^2\chi(\bm k - \bm k') + \mathcal{O}(\Omega_0) \nonumber \\
  \end{eqnarray}
  where the first term comes from diagram $2a$, the second from $2b$, the third from $2c, 2d$ and the last term from $2e$ in Fig. \ref{diag}.  The convolution that enters in the first term is found to be
   (assuming that the FS averages $<\cos 2l_x> = <\cos 2l_y> =0$, where ${\bf l}$ and $-{\bf l}$ are intermediate fermionic momenta in Fig. \ref{diag}a)
   \begin{equation}
(g \star g)(\bm q) =
\frac{1}{2}  \left( \cos{q_x} + \cos{q_y} \right) + \beta^2 \cos{q_x} \cos{q_y}.
\end{equation}
  \begin{figure}
\includegraphics[width=3.5 in]{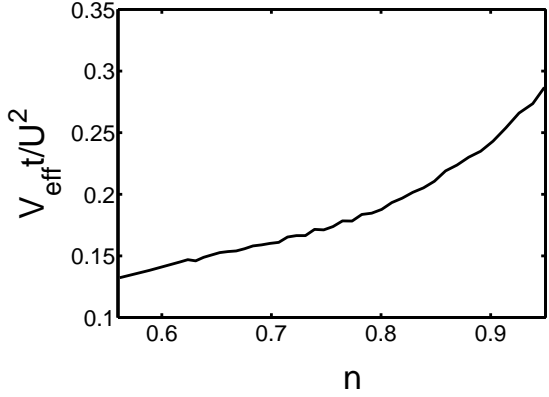}
\caption{The effective interaction $V_{eff} = \lambda/N_F$ for the
g-wave state.  Here, we have chosen $V = 0.5 U$ and $V' = 0.5V$.
All other eigenvalues are subdominant and are not shown here.  }
\label{usimv}
\end{figure}
  Lastly,
  \begin{eqnarray}
  f_3^{(s)}(\bm k, \bm k') &=& aN_F  g(\bm k - \bm k') \ln{\left[ \frac{W}{\Omega_0} \right]} +  \chi_{2}(-\bm k, \bm k') \nonumber \\ &+&  \chi_{2}(\bm k, \bm k') +
     2g(\bm k - \bm k') \chi(\bm k - \bm k') + \mathcal{O}(\Omega_0) \nonumber \\
  \end{eqnarray}
  where $a$ is a constant of order unity.  The first term is obtained from diagram $2a$, the second from $2b$, the third from $2c$, and the last from $2e$ in Fig. \ref{diag}.
 The functions $\chi(\bm q), \chi_{1}(\bm k, \bm k')$, and $ \chi_{2}(\bm k, \bm k')$ are generalized susceptibilities and are all expressible in terms of the one electron Matsubara Green function, $G(p) = (i \omega_p - \epsilon_p )^{-1}$:
 \begin{widetext}
  \begin{eqnarray}
  \chi(\bm q) &=& \int_p G(p)G(p+q),  \  \ \ \ \ \  \ \int_p \equiv \int \frac{d^d p d \omega_p}{(2 \pi)^{d+1} }\nonumber \\
  \chi_{1}(\bm k, \bm k') &=& \int_p g(\bm k - \bm p) g(\bm p - \bm k') G(p)G(p-k-k') \nonumber \\
  \chi_{2}(\bm k, \bm k') &=& \int_p \left[g(\bm p + \bm k) + g(\bm p - \bm k') \right] G(p)G(p+k-k')
  \end{eqnarray}
  \end{widetext}
  Similarly, in the spin-triplet channel, the effective interaction takes the form
     \begin{eqnarray}
 \Gamma^{(t)}(\bm k, \bm k') = && V g(\bm k- \bm k') + U^2 f_1^{(t)}(\bm k, \bm k') + V^2f_2^{(t)}(\bm k, \bm k') \nonumber \\
 && + UV f_3^{(t)}(\bm k, \bm k') + \cdots
 \end{eqnarray}
where
\begin{eqnarray}
f_1^{(t)}(\bm k, \bm k') &=& -\chi(\bm k - \bm k') \nonumber \\
f_2^{(t)}(\bm k, \bm k') &=& f_2^{(s)}(\bm k, \bm k') \nonumber \\
f_3^{(t)}(\bm k, \bm k') &=& -2 g(\bm k - \bm k') \chi(\bm k- \bm k')
\end{eqnarray}
When $\beta = 0$, the contribution from diagram $2a$ disfavors solutions which involve nearest-neighbor pairing.
  For a
  non-zero $\beta$, solutions with second-neighbor pairing amplitude are also suppressed.

In Fig \ref{usimv}, we show the pairing strength of the A$_{2g}$(g-wave) state when $V=0.5U, V'=0.5V$.  We have found that the pairing strengths in other channels are significantly weaker and are therefore not shown in the figure.
   We note
   that,
   interestingly, the  g-wave pairing strength is much greater in the regime $V \sim U$ since it gains energy from the attractive effective interactions that are obtained at $\mathcal{O}(V^2)$, which were neglected in the regime where $ V \sim U^2$.  Thus, while the states with nearest neighbor pairing are severely suppressed
    by momentum-dependence of the first-order terms,  states involving further ranged pairing
     are more enhanced by second-order, KL contributions than they are suppressed by first-order contributions.  
      It is important to stress that in the weak coupling limit, such enhancements cannot be ascribed to a well-defined bosonic `glue', such as those associated with proximate spin or charge density states, since instabilitites toward spin and charge
 ordered phases occur at weak coupling only in exponentially narrow regions in parameter space (close to a perfectly nested Fermi surface, for instance).

\section{Continuum electrons with screened Coulomb interactions}
\label{sec:coulomb}

 The Coulomb interaction is long-ranged, and so never can be safely treated in perturbation theory.  Physically, in a good metal, it should be screened, and so equivalent to an appropriate finite-range model, but because screening at long distances involves electrons with energies arbitrarily close to the Fermi energy, this is something that is subtle to treat in a controlled RG.  Simply to replace the Coulomb interaction with a screened Coulomb interaction, and then to carry out the usual RG from there risks double-counting of certain important physical processes, which are already implicit in the screened form of the interaction.  Thus, to treat the Coulomb problem, we are forced to adopt a more qualitative, although still physically sensible approach, in lieu of the asymptotically exact approach we have pursued up until this section.
 For simplicity, we again consider the explicit case of dilute electrons, in which the dispersion is quadratic.  It is important to note, however, that for small $r_s$, where weak-coupling intuition is most likely to be correct, the screening length is parametrically larger than $k_F^{-1}$, so even the screened interaction is not, in any naive sense,
 short-ranged.~\cite{alexandrov,Chubukov-KaganCM89}

 In 3D, the screened Coulomb interaction
 within the RPA approximation is
 \beq
\tilde U(q) = \frac{4\pi e^2}{q^2 + \kappa^2 \Pi (q)}
\label{u}
\eeq
where
\beq
\Pi (q) = \frac{1}{2} + \frac{4k^2_F - q^2}{8 k_F q} \log{\frac{2k_F +q}{2k_F -q}},
\eeq
 $\kappa = 0.81 k_F r_s^{1/2}$, and $r_s = 1.92 e^2/\hbar v_F$.
 If we calculate the s-wave scattering length and the range of the interaction, as we did in Section \ref{sec:isotropic},
 then $ak_F = me^2 k_F/\kappa^2 \approx 0.8$ and  $r_0 \sim 1/\kappa \sim a/(r_s)^{1/2}$,
so in the ``weakly interacting limit'' $r_s <<1$,  it follows that $r_0 \gg a$.
Because $ak_F= {\cal O}(1)$,
an expansion in powers of the interaction is not
 possible. In addition, the screening of the Coulomb interaction already contains one
 of the diagrams
  (a particle-hole bubble) which for short range interactions participated in the KL renormalization of the irreducible pairing interaction.
  None-the-less, it is still plausible that the same sort of KL calculation (with the screening diagram omitted) will still give a physically reasonable account of the pairing in this case, although the formal justification for this approach is on somewhat less secure footing.

\subsection{ $r_s <<1$}
 For $r_s <<1$,
 since the relevant values of the momentum transfer which enter the various calculations is $|{\bf q}| \sim \kappa << k_F$,
 $\tilde U(q)$ can be well approximated by taking $\Pi (q) \approx \Pi (0) =1$. Then
 \beq
\tilde U(q) = \frac{4\pi e^2}{q^2 + \kappa^2}
\label{u1}
\eeq
 and consequently
\beq
N_F \tilde U(q)
= 0.041 \frac{r_s}{1-\cos{\theta} + 0.164 r_s}
\eeq
where $q=\sqrt{2k_F^2[1-\cos(\theta)]}$.
Expanding this bare $\tilde U(q)$ in angular harmonics one
finds that they are all positive (repulsive),
but that their magnitudes
 decay with increasing $l$, eventually as
 $e^{-l r_s^{1/2}}$ for  $l r_s^{1/2} >>1$ (Ref.\cite{Chubukov-KaganCM89}).

 The KL-type contributions to $\lambda^{(
 l)}$ come from particle-hole renormalizations which involve particle-hole
 bubbles
 (diagrams 2b, 2c and 2d in Fig.\ref{diag}).
For $ l \geq 1$, typical momenta associated with these bubbles
 are of order $k_F$,
 and since
 $N_F \tilde U(k_F) = {\cal O}(r_s)\ll 1$,
 this implies that these terms make
  contributions to $\lambda^{(l)}$
   which are small compared to the scale of the repulsive contributions
  $\sim N_F\tilde U(0)= {\cal O}(1)$.  Obviously then, for
  $l$ not too large, $\lambda^{(
  l)} >0$,  {\it i.e.}, there is no superconductivity.  The situation is different for large $l > 1/r^{1/2}_s$
  owing to the fact that the exchange diagram
  (diagram 2b in Fig.\ref{diag}) has a non-analytic dependence on $\theta$ near $\theta =0$, {\it i.e.} when
  ${\bf k}$ and ${\bf p}$ are nearly parallel.
  Moreover, under these circumstances, the typical internal momenta in  the particle-hole ladder are also nearly parallel to ${\bf k}$ and ${\bf p}$, from which it follows that the relevant interactions
  involve near-zero momentum transfer
  where the interactions are strong, $N_F \tilde U(0) = {\cal O}(1)$ (Ref.\cite{Raghu_11}). Because of
  this small $\theta$ non-analyticity, the
   contribution to $\lambda^{(l)}$ from the exchange diagram scales as $1/l^4$ rather than exponentially in $l$, and so makes the dominant contribution to
   $\lambda^{(l)}$ for large enough $l$.
   This non-analytic contribution is attractive
    independent on the parity of $l$ (Ref.\cite{Kohn-Luttinger}).
    Of course, $T_c$  is always very small for such high $l$ pairing, but,
    these arguments suggest that
    KL-type superconductivity survives.

    \subsection{$r_{wc} > r_s >1$}

   For $r_s \gg 1$, even in the continuum, electrons form a Wigner crystal state, which is manifestly neither a Fermi liquid nor a superconductor.  However, since the critical value, $r_{wc}$, for Wigner crystallization appears to be numerically large, it is reasonable to wonder what happens when $r_s$ is in the range $1 < r_s < r_{wc}$.
 In this range, there is no guarantee that the Fermi liquid fixed point is even a correct starting point for examining the problem - there could be a variety of possible other phases, and, even if one is prepared to ignore all standards of mathematical rigor, there is no controllable expansion because
    $N_F \tilde U(q)$ is of order one for all $q$ which connect points on the FS.

   The approach adopted by AK to address this point was to assume  the same perturbative approach that we have outlined for $r_s \ll 1$ can be applied for $r_s >1$, {\it i.e.}  the pairing vertex is calculated to second order in the screened Coulomb interaction which is taken to have the Yukawa form
   as in Eq.\ref{u1},
    and only  the exchange diagram 2e from Fig. \ref{diag} is evaluated to determine the second-order contribution to the pairing vertex.
      Evaluating $\lambda^{(l)}$ for the first few angular
       harmonics, one then
       finds~\cite{alexandrov} that the interactions
     in the $p$ and $d-$wave channels (among others) are repulsive for $r_s \sim 1$, and only become attractive again
 for unphysicaly large values, {\it i.e.,} $r_s > 54$ for $p-$wave
 and  $r_s > 26.5$ for $d-$wave pairing.  From this AK concluded that the KL mechanism is not a plausible mechanism of superconductivity in systems with Coulomb interactions with $r_s \sim 1$ or greater.

Given that even the starting expression for the screened Coulomb interaction in Eq.  \ref{u} makes little sense in the large $r_s$ limit, where the Thomas Fermi screening length is parametrically smaller than the distance between electrons, it is difficult to judge the validity of this conclusion. We could, with no less  justification, repeat the AK calculation but keeping the full RPA expression for the screened Coulomb interaction, Eq. \ref{u}, rather than the Yukawa form in Eq. \ref{u1}.  In this case, an attractive pairing vertex is obtained already in  first order!
(Second order contributions
 from diagrams 2b-2d in Fig. \ref{diag}
  would
 further enhance the pairing tendencies.)
 As was shown in Ref. \onlinecite{Chubukov-KaganCM89},
  the largest attractive pairing component attractive component is
   $l=1$ for which $\lambda{(l=1)} \approx -0.07$.
 A very similar $\lambda{(l=1)} \approx -0.06$ has been obtained in Ref. \cite{peter} where the effective interactions in spin and charge channels were chosen to fit Monte-Carlo data for compressibility and spin susceptibility.
   In the absence of a justified theoretical computational scheme, it is difficult to gauge which approach is better. It is likely that $p-$wave is the leading attractive pairing component at $r_s >>1$  but at which $r_s$ it prevails and what is $T_c$ are the two issues not settled yet.

 There are other issues in the large $r_s$ limit.
  Due to the proximity of the
 Wigner crystal, the effective interactions are strongly
momentum-dependent.
 Magnetically ordered states are also natural in this limit, and proximity to them can give rise to structure in the pairing vertex related to magnetic fluctuations:  This observation underlies theories of $p-$wave pairing
near a ferromagnetic~\cite{anderson-morel} and
d-wave pairing near an antiferromagnetic
instability~\cite{d-wave}.
However, the quasiparticles themselves may be largely incoherent, 
making it difficult to know precisely what is being paired.
 Even if Fermi liquid theory remains valid, strong forward scattering interactions, which are likely to occur in this limit, can produce large changes in the Fermi liquid parameters, with major consequences for any theory of superconducting pairing.

\section{DMRG solution of extended Hubbard ladders}
\label{sec:ladders}

\begin{figure}
\includegraphics[width=0.5\textwidth]{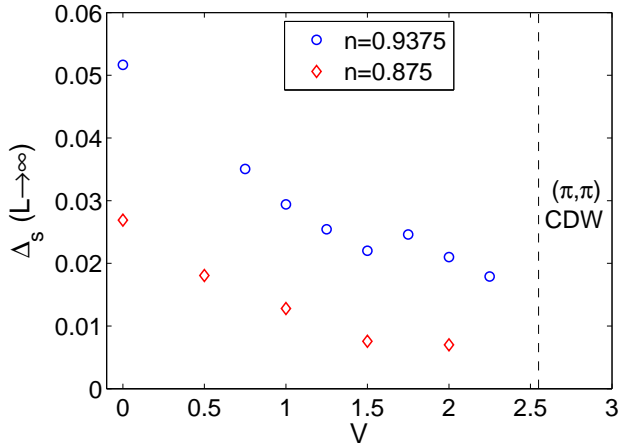}
\caption{The spin gap $\Delta_s$ extrapolated to the thermodynamic
limit for two-leg ladders with electron density $n=0.9375$
(circles) and $n=0.875$ (diamonds) and $t=1$, $U=8$ and $V'=0$, as
a function of $V$. Beyond $V\approx 2.6$, a transition to a state
with pronounced charge density oscillations at a wavevector close
to $(\pi,\pi)$ is observed.}
 \label{fig:Delta_s}
\end{figure}

To go beyond the weak coupling limit, where the energy scales
associated with superconductivity are exponentially small, one has
to resort to numerical methods. In order to address the effect of
extended interactions in the intermediate coupling regime, we have
performed DMRG\cite{White1992} simulations of the extended Hubbard
model (Eq. \ref{su_6}) on ladders of size $L\times 2$. For
$V=V'=0$, this system is
known\cite{Dagotto1992,Dagotto1996,Balents1996,DMRG_ladders} to
have a spin gap in the thermodynamic limit when the electron
density is close to $n=1$. The superconducting correlations are
d-wave like, in the sense that the pair amplitude has an opposite
sign on x and y oriented bonds, and falls off with distance as a
power law which
depends on $n$ 
and $ U/t$. Here, we examine the sensitivity of the
superconducting tendency of the two-leg ladder to adding extended
interactions. We do this by calculating the spin gap and the decay
of a proximity induced superconducting order parameter as a
function of distance, for various values of $V$, $V'$. This study
extends earlier results on the t-J model with a nearest-neighbor
V,\cite{Gazza1999} which found that the pairing survives up to
$V\approx 4J$.

The spin gap is defined as
$\Delta_s=E(S=1)-E(S=0)$ where $E(S)$ is the ground state energy
with spin $S$, extrapolated to the thermodynamic limit
$L\rightarrow\infty$, as a function of $V$. The system sizes used
in the extrapolations where $L=16,32,64$. The spin gap is then
extrapolated to $L\rightarrow\infty$ by fitting $\Delta_s(1/L)$ to
a second order polynomial. The extrapolated value of $\Delta_s$ is
up to a factor of $2$ smaller than $\Delta_s(L=64)$; this
extrapolation is the largest source of error in our results. The
following parameters were used in the calculations: $t=1$, $U=8$,
$V'=0$ and $n=0.875,0.9375$. As $V$ increases, the spin gap
decreases gradually from its $V=0$ value, but remains finite up to
$V\approx 2.5$.

For larger values of $V$, we observe a transition
to a charge density wave (CDW) state, in which there are
pronounced oscillations in the electron density at a wavevector
close to $\vec{Q}=(\pi,\pi)$. We have checked that the CDW  transition
occurs at $V\approx 2.5$ even for the undoped ($n=1$) system. The
doped CDW state supports gapless spin excitations at the edges,
but has a large bulk spin gap $\Delta_s\approx 0.1$. (We infer
this by noticing that in the lowest triplet excitation with
$S_z=1$, $\langle S_z \rangle$ is non-zero only close to the
edges. We have also computed the gap to an excitation with
$S_z=2$, in which $\langle S_z \rangle$ is non-zero in the bulk,
and found that this gap is finite in the thermodynamic limit.)
\begin{figure}
\includegraphics[width=0.5\textwidth]{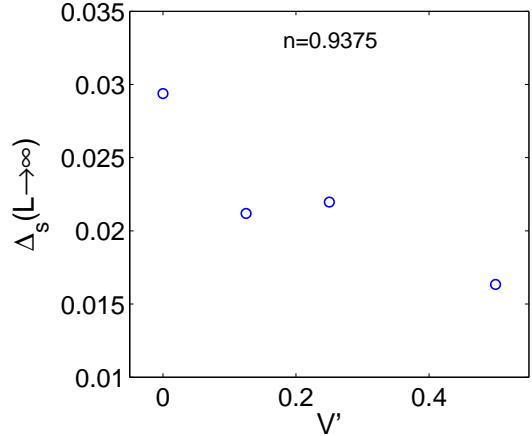}
\caption{The spin gap $\Delta_s(L\rightarrow \infty)$ as a
function of $V'$ for two-leg ladders with $n=0.9375$, $t=1$, $U=8$
and $V=1$.}
 \label{fig:Delta_s_Vp}
\end{figure}

We have also examined the effect of a second-neighbor $V'$ on the
spin gap. Fig. \ref{fig:Delta_s_Vp} shows the spin gap as a
function of $V'$ for $V=1$, $n=0.9375$. Again, we find that while
$\Delta_s$ decreases monotonically upon increasing $V'$, its
effect is not dramatic. E.g., upon reaching $V'=0.5t$, $\Delta_s$
has decreased to about $50\%$ of its $V=1,V'=0$ value.
 \begin{figure}
\includegraphics[width=3.5 in]{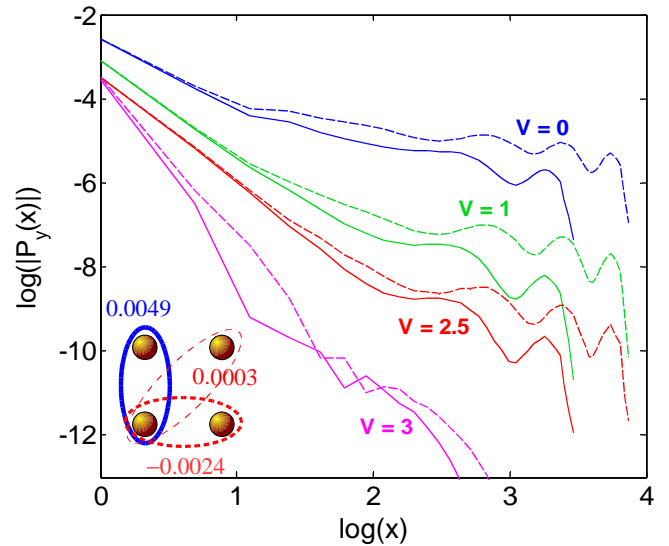}
\caption{Induced superconducting order parameter as a function of
position in a calculation with an edge pair field (Eq.
\ref{eq:edge}) of magnitude $\Delta=0.5$. In these calculations,
$t=1$,$U=8$,$n=0.9375$. Solid (dashed) lines correspond to $L=32$
($L=48$), respectively. Results for $V=0,1,2.5,3$ are shown. The
inset shows the induced order parameter on nearest and
second-nearest neighbor bonds near the middle of an $L=48$ system
with $V=1$.} \label{fig:sc}
\end{figure}

The superconducting response of a ladder system can be
characterized by the rate
 at which an externally induced superconducting
order parameter at the edge decays as we move into the bulk. In a
gapless one-dimensional system, this amplitude decays as a power
law; in a two-leg ladder with a spin gap, this power law can be
shown to be equal to $\frac{1}{4K_c}$,\cite{Berg2009} where $K_c$
is the Luttinger parameter of the (gapless) even charge mode. Fig.
\ref{fig:sc} shows the induced order parameter on a y bond,
\begin{equation}
P_y(x) = \langle \frac{1}{2}\left[c_{\uparrow}(x,1)
c_{\downarrow}(x,2) - c_{\downarrow}(x,1)
c_{\uparrow}(x,2)\right]\rangle,
\end{equation}
as a function of position $x$, on a log-log scale. In this
calculation, the following boundary term was added to the
Hamiltonian:
\begin{equation}
H_{\mathrm{edge}} = \Delta \left[c_{\uparrow}(1,1)
c_{\downarrow}(1,2) - c_{\downarrow}(1,1)
c_{\uparrow}(1,2)+\mathrm{H.c.}\right],\label{eq:edge}
\end{equation}
with $\Delta=0.25$. This term mimics a proximity-induced gap at
the edge due to a nearby bulk superconductor. The results are
shown for systems of length $L=32,48$, density $n=0.9375$, and
$V=0,1,2.5,3$.

The induced superconducting order parameter is seen to decrease
monotonically upon increasing $V$. However, the slope of
$\log(|P_y|)$ vs. $\log(x)$ (the power with which the
superconducting order parameter decays) far away from the edge is
not strongly dependent on $V$, except for $V>2.5$. Although our
systems are not long enough to allow an accurate estimate of the
slope, one can roughly estimate $K_c \sim 0.4-0.6$, well within
the range of divergent superconducting correlations $K_c>0.25$,
and close to the value in which superconducting and CDW
correlations decay with the same exponent, $K_c=0.5$.

The inset of Fig. \ref{fig:sc} shows the induced superconducting
order parameter on various bonds near the middle of the $L=48$
system with $V=1$. As can be seen in the figure, the order
parameter is ``$d_{x^2-y^2}-$ like'', in the sense that the
pairing amplitude on x and y oriented bonds is opposite in sign,
and the amplitude on the diagonal (next-nearest neighbor) is
relatively small. The order parameter has a $d_{x^2-y^2}-$ like
structure for $0<V<V_c \approx 2.5$. For larger values of $V$ (in
the CDW phase), the order parameter has extended s-wave structure,
in which the pairing amplitude has the same phase on x and y
oriented bonds. In this regime, however, the order parameter is
much smaller than for $V<V_c$ and it decays much faster as a
function of distance from the edge (possibly exponentially).

\section{Discussion}
\label{sec:discussion}
In this paper we have shown, using a variety of methods, that unconventional superconductivity arising directly from electron-electron interactions can survive even in the presence of longer (but finite) ranged interactions.
 Firstly, we have shown that the Kohn-Luttinger effect survives even when the range of the interaction exceeds the s-wave scattering length, provided that the range $r_0$ satisfies $r_0 \ll k_F^{-1}$.  In this regime, the leading instability need not occur for an unphysically high angular momentum and in most cases occurs
 for $\ell = 1$.  For lattice electrons near half-filling in the weak-coupling limit, we have found that the $d_{x^2-y^2}$ superconductivity that occurs in the Hubbard model survives in the presence of longer ranged repulsive interactions $V, V', \cdots$, provided that these interactions are no larger than $ \alpha U^2/W$, where $\alpha$ is a constant of order unity.  Lastly, we have shown, using DMRG calculations, that in the intermediate coupling regime, the spin-gap survives even in the presence of a substantial nearest-neighbor interaction, suggesting the stability of $d_{x^2-y^2}$ superconductivity against longer ranged interactions.

Our findings have possible relevance for understanding some aspects of the effect of material specific changes in the electronic structure on the transition temperatures of unconventional superconductors. Although we have shown that unconventional superconductivity from repulsive interactions is a robust phenomenon, which survives in the presence of substantial farther range repulsions, there is none-the-less a strong tendency for such interactions to produce a significant reduction of T$_c$.  Conversely, if screening by a proximate polarizable medium reduces V and VÕ, this could lead to a marked enhancement of T$_c$.

As a point of comparison, recall that in a conventional electron-phonon superconductor, T$_c$ depends on the electron-electron interactions only through $\mu^*$. However, because of retardation, $\mu^* \sim 1/\ln[E_F/\omega_0]$  is largely independent of the bare electron-electron interaction.  Moreover, the effective attraction induced by the phonons is typically highly local, and so is also unlikely to be very sensitive to small changes in the environment.  Thus, a sensitivity to the screening effects of a polarizable environment may be one of the uniquely characteristic features of pairing
 due to electron-electron interaction.

{\bf Acknowledgements}

We thank  A.S. Alexandrov, S. Doniach, T. Geballe,
M. Yu. Kagan and D. J. Scalapino for useful discussions.
 This work was supported in part by  NSF-DMR-0906953 (AVC), DMR-0758356 (SAK), DMR-0757145, DMR-0705472 (EB) and startup funds at Stanford University (SR).
 SR and AVC  also wish to thank the Aspen Center for Physics where part of this work was carried out.

\end{document}